\documentclass[sigconf]{acmart}

\acmSubmissionID{18}
\AtBeginDocument{%
  }

\copyrightyear{2026}
\acmYear{2026}
\setcopyright{cc}
\setcctype{by}
\acmConference[MIG '26]{The 19th ACM SIGGRAPH Conference on Motion, Interaction, and Games}{December 11--13, 2026}{Charleston, SC, USA}
\acmBooktitle{The 19th ACM SIGGRAPH Conference on Motion, Interaction, and Games (MIG '26), December 11--13, 2026, Charleston, SC, USA}
\acmDOI{10.1145/3828647.3857292}
\acmISBN{979-8-4007-2824-2/2026/12}

\usepackage{algorithm}
\usepackage{algorithmic}

\begin{document}

\title{Constructing Dynamic Conversations with Neural Back-channel Generation and Voice-line Scheduling}

\author{Yifang Pan}
\email{evan.pan@mail.utoronto.ca}
\orcid{0000-0001-5194-0359}
\affiliation{%
  \institution{University of Toronto}
  \city{Toronto}
  \country{Canada}
}

\author{Chris Landreth}
\email{c.landreth@rogers.com}
\orcid{0009-0004-5110-2610}
\affiliation{%
  \institution{JALI Research}
  \city{Toronto}
  \country{Canada}
}

\author{Karan Singh}
\email{karansher.singh@utoronto.ca}
\orcid{0000-0002-2232-7480}
\affiliation{%
  \institution{University of Toronto}
  \city{Toronto}
  \country{Canada}
}
\renewcommand{\shortauthors}{Pan, Landreth, and Singh}

\begin{abstract}
We present a novel approach to the problem of combining two voice-line channels of dyadic conversational audio into a unified conversation with controllable back-channeling and turn-taking dynamics.
In voice performances for the game industry, the two tracks of voice-lines are typically recorded or synthesized in isolation. Our method takes two such unsynchronized tracks and uses a neural scheduler to sequence them into a single dynamic conversation, modifying both tracks by inserting or removing pauses and introducing voice overlap as appropriate. We further propose a back-channel predictor that is able to predict opportunity for back-channels for a given conversational turn's audio and content, which the scheduler can then incorporate. We evaluate our scheduler and predictor independently and in combination by: various quantitative comparison; a qualitative and perceptual comparison to prior art and naive baselines; a compelling set of results showing animator control over the diversity, extent and nature of back-channeling and turn dynamics.

\end{abstract}

\ccsdesc[500]{Computing methodologies~Animation}
\ccsdesc[300]{Computing methodologies~Neural networks}

\keywords{speech, audio-driven animation}
\begin{teaserfigure}
  \centering
  \includegraphics[width=\textwidth]{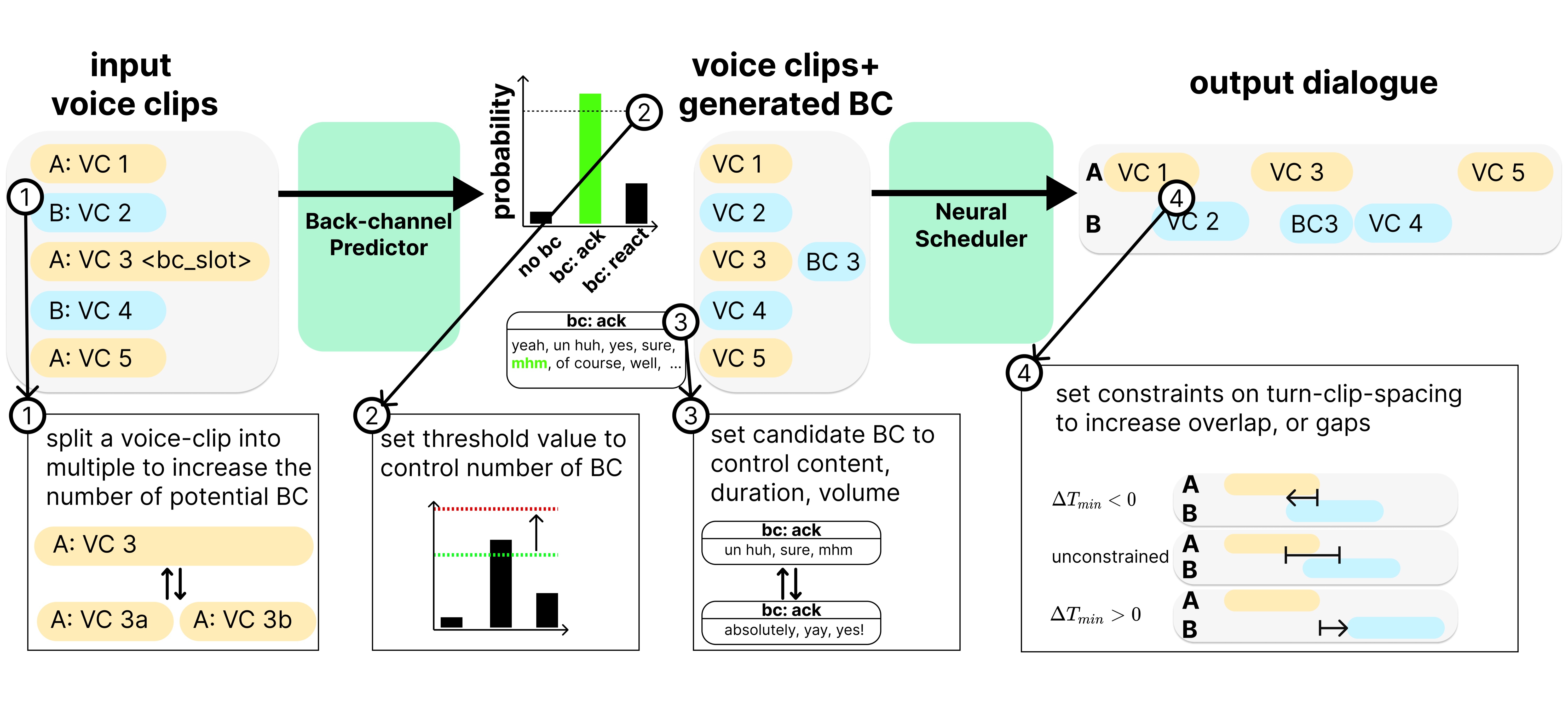}
  \caption{Input turn-clips are processed by the Back-channel Predictor, which determines and synthesizes a candidate verbal back-channel. The Neural Scheduler then sequences the turn-clips and generated back-channels with appropriate temporal gaps or overlap into a conversational dialogue. Circled numbers indicate animator-centric parameters: \textbf{(1)} turn-clip splitting, and \textbf{(2)} a back-channel probability threshold to control back-channel frequency; \textbf{(3)} a back-channel vocabulary with varying valence, duration, and intensity; and \textbf{(4)} timing constraints to control pause and overlap in conversational turn-taking.}
  \Description{Pipeline diagram. Input voice clips from speakers A and B pass through the Back-channel Predictor, which adds a generated back-channel clip, and then the Neural Scheduler, which arranges all clips on a two-channel timeline. Four insets show the animator controls: clip splitting, the back-channel probability threshold, the back-channel vocabulary, and timing constraints.}
  \label{fig:teaser}
\end{teaserfigure}

\maketitle

\section{Introduction}

Voice recognition, synthesis, and {\it vocap} (voice-capture driven character animation) solutions are now commonly used for games, film and virtual avatars \cite{Tan2021NeuralTTS}. A fundamental advantage of voice is scalability, due to its ease and flexibility of recording. Characters in games like Cyberpunk 2077 \cite{edwards2020jalidrivena} are driven by tens of thousands of lines of dialogue recorded in over ten languages. This dialogue however, is typically recorded by voice actors in isolation, read from a script with little preparation or context \cite{backstage}. Even for AAA games, ``approximately 95\% of the time, actors are recorded alone in the booth'' (private communication \cite{pixelaudio}). Consequently, despite expressive individual voice acting, simply splicing these recordings together produces conversational turns that sound uniformly robotic,
lacking the natural overlaps and pauses that characterize interactions.
Further, conspicuously missing in such conversations, are listener interjections or {\em back-channels}.
Back-channels are short verbal utterances like ``mmm-hmmm'', ``cool'' or ``maybe'', that can indicate listener engagement,  enthusiasm, or skepticism, but above all keep the conversation flowing \cite{Clark1991grounding}. While a vocabulary of back-channels can sometimes be recorded separately, and manually edited into conversations by an animator \cite{pixelaudio}, such an approach is tedious and does not scale to increasingly many hours of dialogue to be processed for games like Cyberpunk 2077.

We jointly address the scheduling and back-channel problem with a unified system, and present the first controllable neural solution to producing dynamic conversational turns from a script of independently recorded or TTS (text-to-speech) voice-lines.
As seen in Fig.~\ref{fig:teaser}, our framework operates on a script-ordered set of voice-lines (recorded by voice actors or synthesized with TTS), and produces natural two-speaker dialogue audio via a two-component pipeline. The \textit{Back-channel Predictor} first determines where listener back-channels should occur and of what type, operating at the voice-line level. The \textit{Neural Scheduler} then determines the precise timing of back-channels, turn transitions and voice overlap for both channels jointly. Both components combine learned models with algorithmic inference, and expose hyperparameters that afford fine-grained control over the back-channel frequency, valence, and conversational dynamics of the output at inference time.

Outside of applications in game development, we further see our model as a stepping stone towards more natural and interactive voice LLMs.
Successful text-LLMs like ChatGPT and Gemini increasingly provide voice-LLM support \cite{geminiteam2025geminifamilyhighlycapable}.  These models are trained on synthetic TTS data, and fine-tuned using small conversation datasets (e.g., Fisher or Switchboard \cite{cierifisher}), and thus produce voice assistants that lack back-channeling and natural conversation dynamics.
Our work can transform large text datasets into dynamic conversations using synthesized voices for use in training future conversational voice assistants, and other full-duplex model applications \cite{roypersonaplex}.

\paragraph{Evaluation} We evaluate each component of our model in isolation and the full pipeline end-to-end. The Back-channel Predictor achieves state-of-the-art performance on sentence-level back-channel opportunity prediction (\autoref{sec:bc_classifier}), and the Neural Scheduler produces conversational timing statistics closest to ground truth across nearly all turn-taking metrics (\autoref{tab:turn_taking}). Evaluated end-to-end, our system generates back-channels at the highest frequency while achieving the best temporal alignment with natural listener behavior on Full-Duplex Bench (\autoref{tab:fullduplex}), and is significantly preferred over PersonaPlex \cite{roypersonaplex} in a perceptual user study (\autoref{sec:wholistic_evaluation}).

\paragraph{Contribution} Our principal contribution is, to our knowledge, the first end-to-end pipeline that combines two independently recorded voice-line channels into a unified dyadic conversation with natural, controllable back-channeling and turn-taking dynamics. Its key components are:
\begin{enumerate}
  \item A \textit{Neural Scheduler}, based on our formulation of voice-line scheduling as next-token prediction, in which a full-duplex language model is given the content of every line and learns only its timing.
  \item A \textit{Back-channel Predictor}, built around a multimodal classifier that predicts from a clip's text and audio whether it invites a back-channel, and of what type.
  \item A search-based inference algorithm that uses the scheduler to score candidate timings, exposing animator-centric parameters for back-channel insertion and high-level control over conversational turn dynamics without retraining.
\end{enumerate}

\section{Related Work}
We classify related work as pertaining to text-to-speech synthesis, conversational dynamics, and full-duplex language models.
\subsection{Synthesizing Speech from Text}
TTS has evolved from early sequence-to-sequence models \cite{vandenOord2016wavenet, Wang2017tacotron, Shen2018tacotron2} to efficient non-autoregressive architectures and GAN vocoders \cite{Ren2019fastspeech, Ren2021fastspeech2, Kong2020hifigan, Kim2021vits}. Modern TTS is typically a two-stage process using discrete audio codecs \cite{Zeghidour2022soundstream, Defossez2022encodec}: Large Language Models generate tokens from text \cite{Borsos2022audiolm, Wang2023valle}, which are decoded via diffusion \cite{Borsos2023soundstorm, Shen2023naturalspeech2} or flow matching \cite{Chen2024f5tts, Peng2024voicecraft, Du2024cosyvoice, chatterboxtts2025} into high-fidelity speech \cite{Chen2024valle2, Li2023styletts2}.
Text alone however, is insufficient for dyadic conversation, which requires modeling speaker overlap \cite{Reece2023candor} and listener back-channels \cite{Anderson2025NaturalTurn} absent in text transcripts. While joint-synthesis systems like MoonCast \cite{Ju2025MoonCast} and CoVoMix \cite{Zhang2025CoVoMix2} simulate basic turn-taking, they lack precise control over turn boundaries and do not model back-channels. We address these by using a neural scheduling system to combine single-speaker lines into natural two-channel dialogue, with explicit control over both turn-taking and back-channel dynamics.

\subsection{Controlling Back-channels and Turn Dynamics}
Back-channels provide essential grounding \cite{Clark1989contributing, Clark1991grounding}, where continuers (e.g., \textit{mmmhmmm}) signal turn maintenance \cite{Schegloff1982continuers} and emotional responses (e.g., \textit{oh!}) invite elaboration \cite{Heritage1984changestate, Tolins2014addressee}. Inter-turn timing is equally communicative; while the modal gap for English is around 240~ms \cite{Stivers2009universals}, deviations can signal competition \cite{Trimboli1984switching}, deception \cite{reynolds2011cues}, or deliberation \cite{mcconaghie2021correlations}.
Existing computational models approach these dynamics through multi-class classification \cite{Ekstedt2020TurnGPT, onishi2023prediction, wang2024turntakinga, Lin2025predicting} or continuous Voice Activity Projection (VAP) \cite{Ekstedt2022VoiceAP, Inoue2024RealTimeVAP}. However, these statistical methods struggle with back-channel reliability and lack mechanisms for animator control. We address these limitations by combining statistical and procedural components, enabling explicit control over the timing and content of both turn transitions and back-channels.

\subsection{Full-Duplex Language Models}
The emergence of speech language models has made it possible to train systems that understand spoken input and respond through next-audio-token prediction \cite{xu2025qwen3omnia, xie2024miniomnia, zeng2024glm4voicea}. However, these models process a single audio stream at a time, and are thus unable to generate conversational dynamics such as overlapping speech or back-channels.
Full-duplex models address this by jointly modeling two token streams at all times. dGSLM \cite{Nguyen2022dGSLM} introduced a dual-tower Transformer trained on stereo telephone conversations, while LSLM \cite{ma2024language} and NTPP \cite{wang2025ntppa} extended the paradigm with richer inter-stream interaction. Paired with streaming audio encoders and decoders, systems like Moshi \cite{defossez2024moshi} can process live user speech on one channel while generating responses on the other. However, the scarcity of dual-channel training data results in many of these models producing believable conversational dynamics but incoherent speech. Recent work such as PersonaPlex \cite{roypersonaplex} addresses this through large scale TTS-synthesized training data, but is sparse in back-channeling. We complement this line of research, as our system can generate back-channel-rich dialogue data that could serve as training material for the next generation of full-duplex conversational agents.

\section{Method Overview}
\label{sec:method_overview}
Our framework operates on voice clips, which can either be lines recorded by a voice actor or a synthetic voice generated from text. In conversation, a \textbf{turn} is a contiguous stretch of spoken content by a single speaker that ends when the floor passes to the other. We refer to the voice audio of a turn (or a portion of it) as a voice clip.
Given an ordered set of said voice clips, our system synthesizes natural two-speaker audio via a two-component pipeline (Figure~\ref{fig:teaser}).
The first component, the \textbf{Back-channel Predictor}, determines \textit{where} in the conversation listener back-channels should occur: a coarse, voice-clip classification of whether a back-channel response for the clip is warranted. The second component, the \textbf{Neural Scheduler}, then determines \textit{when} precisely each voice line is played, jointly scheduling both the original input lines and any synthesized back-channels to produce natural turn transitions with voice overlap and back-channels. The pipeline exposes hyperparameters that afford fine-grained control over the conversational dynamics of the output at inference time (Section~\ref{sec:applications}).

\section{Back-channel Predictor}
\label{sec:bc_classifier}
The Back-channel Predictor supplements the input turn clips with synthesized back-channel clips. At its core is a trained classifier that predicts for each voice-clip whether the listener should produce a back-channel and, if so, whether it should be a low-salience acknowledgment or a high-salience reactive response. Predictions are then used to produce a corresponding back-channel clip, which is added to the set of input clips to be scheduled.

We predict back-channels per voice clip rather than continuously over time, as in Voice Activity Projection (VAP)~\cite{Ekstedt2022VoiceAP, Inoue2024RealTimeVAP}. Clips are the unit in which our input is authored, and we found in practice that VAP models predict whether a turn will continue well, but are poor at deciding when to insert a back-channel. The classifier therefore decides only \textit{whether} a clip receives a back-channel, and the Neural Scheduler (\autoref{sec:neural_scheduler}) decides its precise timing.

\subsection{Back-channel Prediction Classifier}
\label{sec:bc_labels}
Given a voice-clip, the classifier measures the semantic and auditory opportunity for a back-channel response, and if so, what type. We organize labels by \textit{emotional salience} rather than functional or linguistic taxonomy, as back-channel choices are highly personal, depending on speaker intent ~\cite{morikawa2022determining}, and varying with speaker-listener relationship~\cite{fukunaga2025backchannel}. This yields three classes:

\begin{itemize}
    \item \textbf{No-back-channel} (class~0): the clip presents no opportunity for a back-channel response.
    \item \textbf{Acknowledgment} (class~1): the clip invites a neutral, low-salience response (e.g.,~\textit{uh-huh}, \textit{yeah}).
    \item \textbf{Reactive} (class~2): the clip warrants an emotional, high-salience response (e.g.,~\textit{wow, really?}, \textit{for sure, absolutely!}).
\end{itemize}

This salience-based organization directly governs the behavioral impact of each back-channel on the perceived dynamics of the conversation. Once classified, a back-channel can be produced by sampling a voice line from a user-defined pool of back-channel audio, or generated through TTS in the same voice from a user-defined back-channel vocabulary.

\subsection{Input Representation}
\label{sec:bc_input}
The classifier is conditioned on a windowed, linearized view of the conversation, centered on a candidate back-channel site. For a target voice clip at index~$t$, the input is constructed as follows. A context window of $\pm k$ clips (we use $k = 5$) around~$t$ is extracted. Each clip within this window is prefixed with a speaker-relative tag: the speaker of the target clip is always labeled \texttt{A}, and the other speaker \texttt{B}, so the model learns speaker-role-invariant representations. Clips are delimited by \texttt{</s>} boundary tokens, and a special \texttt{<bc\_slot>} token is appended to the target clip to mark the candidate insertion site. The classifier predicts the back-channel class at this marked position. Since prediction operates at the clip level, the density of candidate sites, and therefore the number of back-channels inserted into a conversation, can be controlled by varying the granularity at which longer turns are segmented into clips.

Earlier back-channel decisions are fed back into the context: a clip that received a back-channel is followed by its class as a tag token (e.g.\ \texttt{[BC:Acknowledgment]}), which lets the classifier account for recent back-channel density. An input might look like:
\begin{quote}
\small\texttt{A: I was at Walmart today [BC:Acknowledgment] </s> \\
A: Shopping for some clothes </s> \\
B: I was there too </s> \\
A: I didn't see you there. <bc\_slot> </s> \\
B: We probably went at different times </s>}
\end{quote}
The target speaker's waveform for the clip (capped at 15\,s) is paired with the text input, providing prosodic cues.

\subsection{Training Data}
\label{sec:bc_data_prep}
We source training examples from the natural split of the Seamless Interaction dataset, which provides dual-channel audio with sentence-level transcripts~\cite{seamless}. Back-channel annotations are derived automatically from text using a two-stage pipeline. First, a rule-based pre-filter identifies candidate clips by requiring that each be short (under 1.0s or at most 3 words), temporally isolated from the same speaker's surrounding utterances (at least 0.3s silence on both sides), and preceded by the other speaker within 2 seconds. Candidates passing this filter are then submitted to a Qwen LLM, which classifies each into one of five back-channel types, Neutral, Positive, Emotional, Anticipation, and Confirmation, as defined in prior work~\cite{morikawa2022determining},  or rejects it as non-back-channel, using a $\pm$10s conversational context window. These types are then collapsed into a 3-class scheme for training: \textit{No-back-channel}, \textit{Acknowledgment} (if Neutral or Positive), and \textit{Reactive} (Emotional, Anticipation, or Confirmation). The pipeline yields 33,977 Reactive and 137,808 Acknowledgment back-channel clips. We additionally sample negative examples at a 5:1 ratio relative to positive instances, yielding 703,140 No-back-channel clips. The dataset is split 96/2/2 for train, validation, and test.

\subsection{Model Architecture and Training}
\label{sec:bc_model}
The classifier uses a late-fusion architecture combining a text encoder with a frozen speech encoder. The text branch is a fine-tuned RoBERTa-base encoder~\cite{liu2019roberta} extended with back-channel-specific special tokens. The audio branch uses a frozen HuBERT-base encoder~\cite{hsu2021huberta} with a learnable weighted sum over all 13 hidden states, following the SUPERB protocol~\cite{yang2021superb}, downsampled via a stride-4 convolution to 12.5\,Hz. A learned fusion CLS token is prepended to the concatenated text and audio representations, augmented with modality embeddings, and processed by a 2-layer Transformer encoder. The CLS output is passed through a linear head projecting to $C{=}3$ classes. Full architecture details are available in our code repository.\footnote{\url{https://github.com/EvanSamaa/NeuralSchedularRelease}}

The model is trained with cross-entropy loss using inverse-square-root class weighting to mitigate the imbalance between \texttt{No-back-channel} and the two back-channel classes. HuBERT parameters remain frozen throughout; all other parameters are optimized jointly. Training proceeds for up to 10 epochs with early stopping on validation macro-F1 (patience of 3 epochs).

\subsection{Classifier Evaluation}
We compare our model against the state-of-the-art back-channel classifiers of \citeauthor{wang2024turntakinga} and \citeauthor{Lin2025predicting}\footnote{We found \citeauthor{Lin2025predicting}'s original training data contains significant leakage of training labels in their input, resulting in inflated performance. Our reimplementation of their released architecture on our data yields performance more consistent with prior work.}, along with ablated variants of our own model. Both \citeauthor{wang2024turntakinga} and \citeauthor{Lin2025predicting} are originally formulated as tri-class classifiers over \{continue, switch, back-channel\}; for a fair comparison we retrained both for the binary task \{Back-channel, No-back-channel\}, and we ensure all models receive information regarding future context. All models are trained on our processed dataset (\autoref{sec:bc_data_prep}) and evaluated on held-out data.

We report five metrics, all computed on the binary BC-vs-NoBC decision: \textbf{P} precision and \textbf{R} recall, \textbf{F1}, the harmonic mean of precision and recall at the model's argmax decision; \textbf{AUC}, the area under the ROC curve, which measures how well the model ranks examples by back-channel likelihood independently of any threshold; and \textbf{EER}, the equal-error-rate operating point on the ROC curve where false-positive and false-negative rates coincide ($\mathrm{EER} \approx 1 - \mathrm{AUC}$ for well-behaved curves). Higher F1 and AUC are better; lower EER is better. AUC and EER are derived from the calibrated score $1 - P(\text{No-back-channel})$, so they reflect the full probability output rather than the hardened prediction.

We additionally study four ablations, trained without corresponding input contexts: \textbf{no audio} silences the waveform; \textbf{no future text} drops context turns following the target; \textbf{no past text} drops context turns preceding the target; \textbf{no current text} removes the words on the target turn itself, leaving only its turn marker.

\autoref{tab:bc_classifier} reports results on the held-out test set. Our full model attains an AUC of $0.798$, outperforming both prior baselines. Among the ablations, removing audio causes the largest recall drop ($0.49 \!\to\! 0.25$) with little effect on precision, indicating audio is primarily responsible for surfacing back-channel-eligible moments. Removing future or current text each costs $0.07$--$0.11$ AUC, confirming that linguistic content in and immediately following the target turn carries substantial signal. By contrast, removing past context barely affects performance ($\Delta\text{AUC} \le 0.03$), suggesting the model relies far more on the current turn and its immediate continuation than on broader conversational history.

The classifier's reliance on audio also suggests why the LLM annotator used to label our training data (\autoref{sec:bc_data_prep}) is not a substitute for the classifier: it works from text alone, and it labels back-channels that a listener has already produced, whereas at inference time no listener response exists.
\begin{table}[t]
\centering
\small
\caption{Binary back-channel classification results on held-out data. Higher F1, P, R, AUC are better; lower EER is better.}
\label{tab:bc_classifier}
\begin{tabular}{lccccc}
\toprule
Model & F1 $\uparrow$ & P $\uparrow$ & R $\uparrow$ & AUC $\uparrow$ & EER $\downarrow$ \\
\midrule
\multicolumn{6}{l}{\textit{Baselines}} \\
\citeauthor{wang2024turntakinga}     & 0.43     & 0.46     & 0.40     & 0.73     & 0.33 \\
\citeauthor{Lin2025predicting}       & 0.42     & 0.34     & 0.55     & 0.69     & 0.36 \\
\midrule
\multicolumn{6}{l}{\textit{Ours}} \\
Full model                           & \textbf{0.51} & 0.53 & \textbf{0.49} & \textbf{0.80} & \textbf{0.28} \\
\quad w/o audio                      & 0.34 & \textbf{0.55} & 0.25 & 0.75 & 0.32 \\
\quad w/o future text                & 0.38 & 0.36 & 0.40 & 0.69 & 0.37 \\
\quad w/o current text               & 0.43 & 0.41 & 0.46 & 0.73 & 0.33 \\
\quad w/o past text                  & 0.48 & 0.52 & 0.45 & 0.77 & 0.30 \\
\bottomrule
\end{tabular}
\Description{Table of F1, precision, recall, AUC and EER for two baseline classifiers and for our full model with four ablations. Our full model has the best F1, recall, AUC and EER.}
\end{table}
\begin{table}[t]
\centering
\small
\caption{Per-class performance of our full model on the three-way merged task (No-back-channel / Acknowledgment / Reactive).}
\label{tab:bc_per_class}
\begin{tabular}{lcccccr}
\toprule
Class & F1 $\uparrow$ & P $\uparrow$ & R $\uparrow$ & AUC $\uparrow$ & EER $\downarrow$ & Support \\
\midrule
No-back-channel  & 0.89 & 0.88 & 0.90 & 0.80 & 0.28 & 8{,}891 \\
Acknowledgment & 0.44 & 0.46 & 0.43 & 0.79 & 0.28 & 1{,}709 \\
Reactive       & 0.23 & 0.24 & 0.22 & 0.76 & 0.31 & 400   \\
\bottomrule
\end{tabular}
\Description{Table of F1, precision, recall, AUC, EER and support for each of the three classes. No-back-channel scores highest and Reactive lowest.}
\end{table}

The tri-class breakdown (\autoref{tab:bc_per_class}) shows the model handles No-back-channel well ($F1 = 0.888$) but struggles to separate the two BC subtypes. Acknowledgment retains an AUC ($0.792$) on par with the binary task, while Reactive is substantially harder ($F1 = 0.229$), reflecting both its lower frequency (400 examples) and greater linguistic and prosodic variability. This aligns with \citet{onishi2023prediction}, who similarly find that fine-grained back-channel categories are inherently harder to predict. The inherent ambiguity of back-channel type motivates our decision to expose animator control as an explicit stage in the pipeline.

\section{Neural Scheduler}
\label{sec:neural_scheduler}
While the Back-channel Predictor augments input voice-clips with back-channel clips, the Neural Scheduler orders the timing of all clips.
The core of the Neural Scheduler is a full-duplex LLM fine-tuned on the task of scheduling an ordered set of voice lines. This section formally introduces the formulation of voice-clip scheduling, discusses how the model is trained, inference strategies, as well as evaluations of the scheduler.

\subsection{Scheduling as Next Token Prediction}
\label{sec:scheduler_formulation}
\begin{figure}
    \centering
    \includegraphics[width=\linewidth]{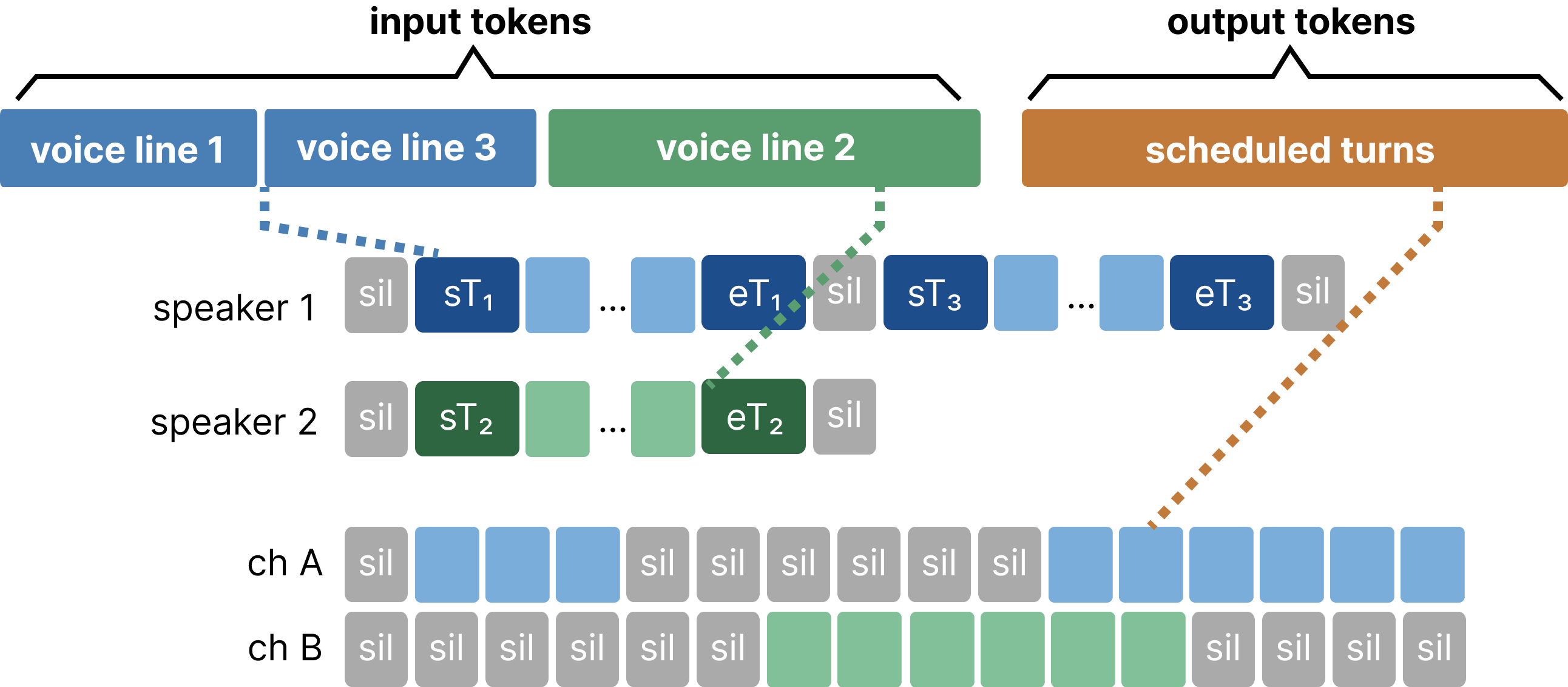}
    \caption{Illustration of how voice-line scheduling can be formulated as a next-token prediction problem.}
    \label{fig:token_organization}
\Description{Diagram of the token layout. The input region holds each speaker's voice lines as separate token sequences with start and end turn markers and silence separators; the output region holds two frame-aligned channels, A and B, in which speech and silence tokens encode the timing.}
\end{figure}
We formulate turn scheduling as a next token prediction problem by introducing two structurally distinct token regions: an \textit{input} region that encodes the input turn-clips of each speaker, and an \textit{output} region where the model generates interleaved output tokens, as shown in \autoref{fig:token_organization}.
\paragraph{Input}
Each speaker's input is an independent, non-interleaved sequence of audio tokens. For speaker~$s \in \{A, B\}$, the input $\mathbf{s}^{(s)}$ is a flat sequence of discrete audio tokens structured into $N_s$ clips by per-turn start and end marker tokens \texttt{[sT}$_n$\texttt{]} and \texttt{[eT}$_n$\texttt{]}, with a silence token $\text{sil}$ separating consecutive clips:
\begin{equation}
    \mathbf{s}^{(s)} = \bigl(\text{sil};\; \texttt{[sT}_n\texttt{]};\; \mathbf{c}^{(s)}_n;\; \texttt{[eT}_n\texttt{]}\bigr)_{n=1}^{N_s}
\end{equation}
where $\mathbf{c}^{(s)}_n$ denotes the speech tokens for clip $n$ of speaker $s$, as illustrated in \autoref{fig:token_organization}. Each clip's speech tokens are drawn from an audio tokenizer; the single silence token $\text{sil}$ between clips is a structural separator only, as it carries no temporal information and does not encode the actual duration of the pause, only that a clip boundary exists. As a result, the inputs are purely \textit{content} representations: they specify what each speaker says and in what order, but are entirely stripped of timing. The two inputs generally have different lengths, reflecting the different number and duration of each speaker's clips. The full input region is formed by concatenating both inputs sequentially, delimited by per-speaker boundary markers \texttt{[sA]}, \texttt{[eA]}, \texttt{[sB]}, \texttt{[eB]}:
\begin{equation}
    \mathbf{x}_{\text{in}} = (\texttt{[sA]};\; \mathbf{s}^{(A)};\; \texttt{[eA]};\; \texttt{[sB]};\; \mathbf{s}^{(B)};\; \texttt{[eB]})
\end{equation}
\paragraph{Output}
The output region $\mathbf{x}_{\text{out}}$ is where timing lives. It is a strictly interleaved, frame-aligned sequence of audio tokens:
\begin{equation}
    \mathbf{x}_{\text{out}} = (a^{(A)}_1,\; a^{(B)}_1,\; a^{(A)}_2,\; a^{(B)}_2,\; \ldots,\; a^{(A)}_T,\; a^{(B)}_T)
\end{equation}
of length $2T$, where each pair $(a^{(A)}_t, a^{(B)}_t)$ represents one frame (at 12.5~Hz) of both channels simultaneously, similar to the formulation in \cite{wang2025ntppa}. Frames with no active speech are assigned the silence token $\text{sil}$, so the output region encodes the full temporal structure of the conversation including pauses, overlaps, turn transitions, through the pattern of speech and silence tokens across both channels.

The Neural Scheduler therefore learns the conditional distribution $p_\theta(\mathbf{x}_{\text{out}} \mid \mathbf{x}_{\text{in}})$, where the input region supplies the \textit{what} and the output region captures the \textit{when}. Crucially, $\mathbf{x}_{\text{out}}$ contains the same speech tokens as $\mathbf{x}_{\text{in}}$, but interleaved across both channels with silence tokens encoding the pauses, overlaps, and turn transitions that constitute conversational dynamics.

\subsection{Training Objective}
\label{sec:scheduler_training}

Standard next-token prediction (NTP) loss alone is insufficient for training the Neural Scheduler, which we demonstrate in~\autoref{sec:ablation}. The fundamental difficulty is a severe class imbalance inherent to the output token sequence $\mathbf{x}_{\text{out}}$.
Within a speech turn, the task of predicting the next token can be trivially solved by simply tracking the token indices in $\mathbf{x}_{\text{in}}$.
Structural transition events, turn changes, mutual pauses, and simultaneous speech are collectively rare relative to within-turn continuation steps.
As a result, a model that consistently predicts ``continue the current speaker'' achieves deceptively low cross-entropy while failing to capture the very events that define conversational timing dynamics. We call this \textit{inertia bias}: since same-token continuations dominate the training signal, the model gravitates toward repeating the current channel's token and never acquires a reliable internal representation of when a transition is imminent.

To counteract inertia bias, the Neural Scheduler is trained with a composite objective. The primary loss is a standard NTP cross-entropy over the $2T$ tokens of the output region:
\begin{equation}
\mathcal{L}_{\text{NTP}} = -\frac{1}{2T}\sum_{t=1}^{T}\sum_{s\in\{A,B\}}
  \log p_\theta\!\left(a^{(s)}_t \;\middle|\; \mathbf{x}_{\text{in}},\, a^{(A)}_{<t},\, a^{(B)}_{<t}\right)
\end{equation}
The auxiliary duration head is a two-layer MLP from the transformer's final hidden state to 32 classes (\autoref{sec:scheduler_architecture}), trained to predict the \textit{remaining silence duration} at each position: how many more consecutive frames the current token will persist on that channel before a transition. Concretely, the ground-truth label at position $(t, s)$ is
\begin{equation}
d^{(s)}_t =
\begin{cases}
  0                                  & \text{if } a^{(s)}_t \text{ is a speech token,}\\
  \text{min }\!\bigl(r^{(s)}_t,\; 31\bigr) & \text{if } a^{(s)}_t = \text{sil},
\end{cases}
\end{equation}
where $r^{(s)}_t$ is the countdown of consecutive silence frames remaining at frame $t$ on channel~$s$, including the current frame. Class~0 indicates that the channel is producing speech (no silence stretch to measure); classes~1--31 encode the remaining silence duration in 80\,ms increments, with class~31 acting as a catch-all for silences of 31 or more frames (${\geq}2.48$\,s). This labeling injects gradient signal at every frame \emph{inside} a silence stretch, compelling the model to build an explicit, future-facing estimate of when the current segment will end rather than reasoning myopically about the next single frame. The duration cross-entropy is
\begin{equation}
\mathcal{L}_{\text{dur}} = \mathrm{CE}\!\left(h_\theta(\mathbf{x}_{\text{out}}),\, \mathbf{d}\right)
\end{equation}
The composite training objective is then a weighted sum of the two, in all of our experiments we use $\lambda_{\text{dur}} = 0.5$
\begin{equation}
\mathcal{L} = \mathcal{L}_{\text{NTP}} + \lambda_{\text{dur}}\,\mathcal{L}_{\text{dur}}
\end{equation}

\subsection{Architecture}
\label{sec:scheduler_architecture}

The Neural Scheduler must process both single-channel script tokens in $\mathbf{x}_{\text{in}}$ and dual-channel interleaved tokens in $\mathbf{x}_{\text{out}}$. Rather than employing separate models per channel as in Moshi \cite{defossez2024moshi}, we use a single decoder-only LLM that handles both regions, following the interleaved-output paradigm of NTPP \cite{wang2025ntppa} and SyncLLM \cite{veluri2024turnbased}. The backbone is GLM-4-Voice \cite{zeng2024glm4voicea}, which shares its architecture with the Qwen family \cite{xu2025qwen3omnia}.
We make three modifications to adapt the base model to the scheduling task.

\begin{figure}
    \centering
    \includegraphics[width=1\linewidth]{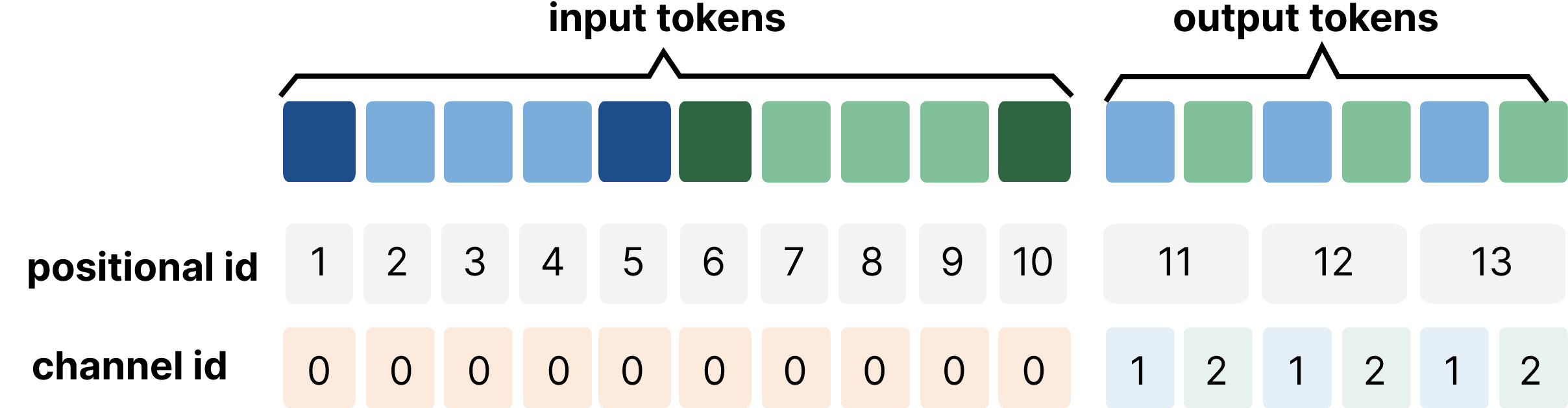}
    \caption{Illustration of positional id and channel id assignment for the corresponding embeddings.}
    \label{fig:positional_id}
\Description{A token sequence with two rows of numbers beneath it. Input tokens have positional ids 1 to 10 and channel id 0; each pair of output tokens shares one positional id (11, 12, 13) and has channel ids 1 and 2.}
\end{figure}

\paragraph{Channel embeddings}
To help the model disambiguate between token regions and between speakers, we introduce a learned channel embedding $\mathbf{e}_c \in \mathbb{R}^{d}$ for each of three channel identifiers $c \in \{0, 1, 2\}$. Tokens in the input region $\mathbf{x}_{\text{in}}$ receive the embedding $\mathbf{e}_0$; tokens on channel~A in the output region $\mathbf{x}_{\text{out}}$ receive $\mathbf{e}_1$; tokens on channel~B receive $\mathbf{e}_2$ (as shown in \autoref{fig:positional_id}). These embeddings are added to the query and key representations in each self-attention layer, enabling the model to distinguish between the single-channel input format and the two interleaved output channels throughout the network.

\paragraph{Positional embedding}
We modify the rotary position encoding to reflect the temporal structure of each region. In the input region $\mathbf{x}_{\text{in}}$, positions increment sequentially by one per token. In the output region $\mathbf{x}_{\text{out}}$, the two tokens at each frame $(a^{(A)}_t, a^{(B)}_t)$ share the same position index, so that the model's notion of temporal order is consistent with the frame-aligned structure: position indices increment by one per frame rather than per token. Concretely, for a token at absolute index $i$ in the output region with dialogue offset $\delta = i - |\mathbf{x}_{\text{in}}|$, the position index is $|\mathbf{x}_{\text{in}}| + \lfloor \delta / 2 \rfloor$.

\paragraph{Duration prediction head and vocabulary}
We extend the model with the duration prediction head described in \autoref{sec:scheduler_training}: a two-layer MLP projecting from the hidden dimension to 32 duration classes (corresponding to countdown values $0, 1, \ldots, 30$ and a catch-all class $31$ for durations $\geq 31$ frames). Additionally, we expand the vocabulary with structural boundary tokens, per-speaker script delimiters (\texttt{[sA]}, \texttt{[eA]}, \texttt{[sB]}, \texttt{[eB]}), dialogue delimiters (\texttt{[sD]}, \texttt{[eD]}), and per-turn markers (\texttt{[sT}$_n$\texttt{]}, \texttt{[eT}$_n$\texttt{]}), to demarcate the regions and turns described in \autoref{sec:scheduler_formulation}.

\subsection{Dataset Processing}
\label{sec:scheduler_data}

We train the Neural Scheduler on the audio data from the Seamless Interaction dataset~\cite{seamless}, a large-scale collection of two-channel conversations recorded on separate microphone channels. Processing each recording into a training example proceeds through four stages: audio tokenization, input and output region construction, turn ID assignment, and duration labeling. We deliberately exclude the Fisher corpus~\cite{cierifisher} from training, reserving it exclusively for evaluation to avoid train-test contamination and to enable fair comparison against prior works.
\paragraph{Audio tokenization}
Each channel is independently encoded into a sequence of discrete speech tokens using the WhisperVQ encoder from GLM-4-Voice~\cite{zeng2024glm4voicea}. A voice activity detection (VAD) model~\cite{SileroVAD} is then applied to each channel to obtain per-channel speech timestamps. Non-speech frames identified by VAD are replaced with the silence token $\text{sil}$, yielding two token streams that faithfully represent both the acoustic content and the activity pattern of each speaker.
\paragraph{Input and output region construction}
The input region $\mathbf{x}_{\text{in}}$ is constructed by encoding each channel's speech content into a compact, non-interleaved script. Consecutive runs of silence tokens are collapsed into a single $\text{sil}$, retaining only the turn boundaries rather than the raw temporal gaps. Each turn's content is delimited by start and end markers $\texttt{[sT}_n\texttt{]}$ and $\texttt{[eT}_n\texttt{]}$, with a single silence token separating consecutive turns. For turns containing internal pauses (e.g., mid-utterance hesitations), a single $\text{sil}$ is inserted between the sub-segments within the turn markers as well. The two per-channel scripts are concatenated sequentially and wrapped by per-speaker boundary markers as described in~\autoref{sec:scheduler_formulation}. The output region $\mathbf{x}_{\text{out}}$ is formed by interleaving the two silenced token streams at each frame, preserving the full temporal structure including pauses and overlaps. Long recordings are segmented into windows of at most 1\,024 tokens (approximately 40\,s of conversation at 12.5\,Hz per channel). During training, the cross-entropy loss is applied only over the output region; the input prefix is masked.
\paragraph{Turn ID assignment}
Turn IDs annotate the input region to delineate conversational turns across both speakers. To assign them, VAD-derived speech segments from both channels are jointly sorted by onset time, and temporally adjacent segments from the same speaker are sometimes ($p=50\%$) merged into contiguous turns. A 50\,\% overlap rule then handles back-channels: if at least half of a segment's duration overlaps with the temporal span of an already-assigned turn, the segment is absorbed into that turn rather than initiating a new one, meaning that back-channels share the same turn ID as the active speech turn. Turn IDs are renumbered contiguously after all merging is complete, producing the indices $n$ used in the turn markers $\texttt{[sT}_n\texttt{]}$ and $\texttt{[eT}_n\texttt{]}$.
\paragraph{Remaining duration labels}
Each position in the output region is annotated with a remaining duration label as described in~\autoref{sec:scheduler_training}. These labels provide per-frame supervision for the auxiliary duration prediction head.

\subsection{Evaluation of Neural Scheduler}
We evaluate the Neural Scheduler through an ablation of the duration prediction objective (\autoref{sec:ablation}) and a quantitative comparison of conversational dynamics against baselines (\autoref{sec:quantitative}). Qualitative examples of its schedules (\autoref{sec:qualitative}) and the evaluation of the full system are presented in \autoref{sec:wholistic_evaluation}.

\subsubsection{Ablation: Duration Prediction Objective}

\label{sec:ablation}
\begin{figure}
    \centering
    \includegraphics[width=\linewidth]{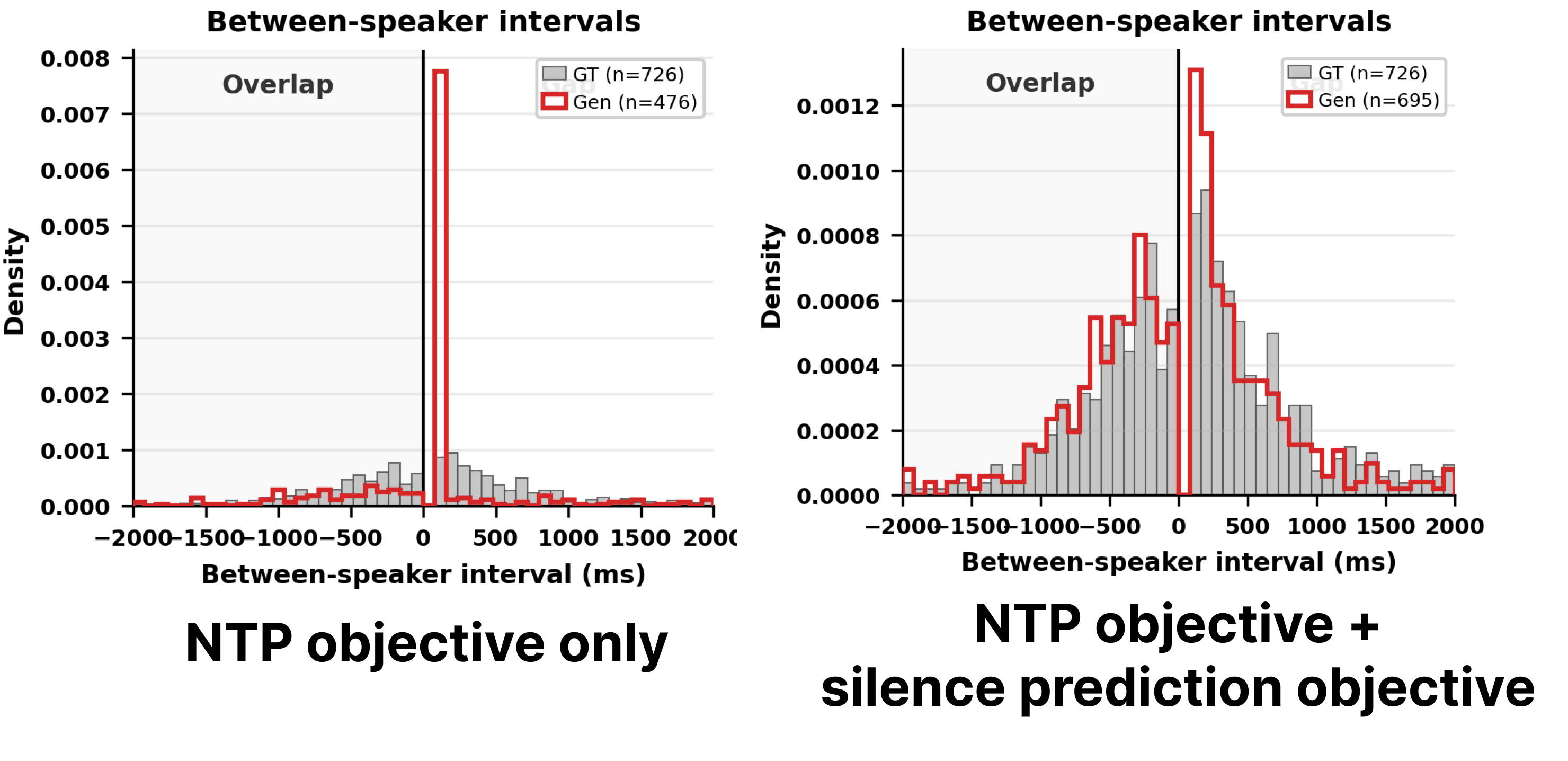}
    \caption{Between-speaker intervals (negative: overlap; positive: gap) in generated vs.\ ground-truth dialogue. Without duration prediction (left), generated intervals collapse into a spike near zero; with it (right), they track ground truth.}
    \label{fig:silence_ablation}
\Description{Two histograms of between-speaker intervals from -2000 to 2000 ms, comparing generated and ground-truth dialogue. With the next-token objective only, generated intervals collapse into a single spike just above zero; with the added silence prediction objective, the generated distribution closely follows the ground truth.}
\end{figure}
\autoref{fig:silence_ablation} ablates the auxiliary duration prediction objective described in \autoref{sec:scheduler_training}. Without it (left), inertia bias causes the generated between-speaker intervals to collapse into a narrow spike just above zero, as the model learns to perpetually continue the current token rather than anticipate transitions. Adding duration prediction (right) resolves this collapse, confirming that the auxiliary objective successfully counteracts inertia bias. Notably, the resulting distribution closely tracks ground truth across both overlaps and gaps, suggesting that the model has learned realistic turn timing dynamics.

\subsubsection{Quantitative Results}
\label{sec:quantitative}
We quantitatively evaluate our Neural Scheduler's understanding of conversational dynamics following the established protocol of NTPP~\cite{wang2025ntppa} and Moshi~\cite{defossez2024moshi}, where models are prompted with 10 seconds of audio and tasked to complete the conversation on unseen Fisher corpus~\cite{cierifisher} samples. Performance is measured by comparing corpus statistics against ground truth across four VAD-derived turn-taking metrics: an \textbf{IPU} (Inter-Pausal Unit) is a contiguous speech segment on a single channel; a \textbf{Pause} is a within-speaker silence where the same speaker resumes; a \textbf{Gap} is a between-speaker silence followed by a turn change; and an \textbf{Overlap} is any interval of simultaneous activity on both channels. For each metric, we report the mean absolute difference $|\Delta|$ in cumulated duration (seconds per minute) against ground truth, where lower values indicate closer alignment with natural conversational dynamics.

We compare against Moshi~\cite{defossez2024moshi} at temperature (0.8, 0.9, 1.0), NTPP~\cite{wang2025ntppa}, dGSLM~\cite{Nguyen2022dGSLM}, LSLM~\cite{ma2024language}, and the naive cascade baseline from~\cite{wang2025ntppa}, with baseline numbers taken directly from the respective publications under the same evaluation conditions. All models are evaluated on the same held-out test split of the Fisher corpus~\cite{cierifisher}.

\begin{table}
\centering
\small
\caption{Turn-taking statistics (cumulated duration/min) for
generated dialogues, compared to ground truth using mean absolute differences ($|\Delta|$) between predicted and ground truth values. Lower $|\Delta|$ values indicate better performance, with the best results highlighted in bold.}
\label{tab:turn_taking}
\begin{tabular}{lcccc}
\toprule
 & \multicolumn{4}{c}{\textbf{Cumulated duration /min}} \\
\cmidrule(lr){2-5}
\textbf{Model} & $|\Delta\text{IPU}|$ & $|\Delta\text{Pause}|$ &
$|\Delta\text{Gap}|$ & $|\Delta\text{Overlap}|$ \\
\midrule
Naive Cascade         & 4.3  & 5.5 & 0.9 & 3.6 \\
dGSLM                 & 4.6  & 3.6 & 1.8 & 1.9 \\
LSLM                  & 4.1  & 3.4 & 1.5 & 2.3 \\
NTPP & 3.3  & 2.8 & 1.4 & 1.9 \\
Moshi$_{0.8}$         & 16.0 & 6.8 & 8.3 & 2.1 \\
Moshi$_{0.9}$         & 6.4  & 2.7 & 3.3 & 1.1 \\
Moshi$_{1.0}$         &  \underline{0.3}  &  \underline{0.6} & \textbf{0.3} &  \underline{0.8} \\
\midrule
\textbf{Neural Scheduler}      & \textbf{0.1} & \textbf{0.1} &  \underline{0.6} & \textbf{0.4} \\
\bottomrule
\end{tabular}
\Description{Table of absolute differences from ground truth in IPU, pause, gap and overlap duration per minute for eight models. The Neural Scheduler has the smallest difference on three of the four metrics.}
\end{table}

Table~\ref{tab:turn_taking} shows that our Neural Scheduler produces conversational dynamics closest to ground truth across nearly all metrics, outperforming both cascaded and full-duplex baselines. This is consistent with the nature of our approach: rather than generating content and timing jointly, the scheduler operates on pre-synthesized voice lines with known linguistic content, leaving timing as the sole degree of freedom to optimize.

Our formulation also sidesteps the intelligibility degradation observed in existing full-duplex models like Moshi~\cite{defossez2024moshi}, which relies on softmax temperature to trade off turn-taking naturalness against speech quality during token inference, higher temperatures improve timing behavior at the cost of intelligibility. Because our model schedules fixed voice lines rather than generating tokens, it is immune to this tradeoff entirely.

\subsection{Inference}
\label{sec:scheduler_inference}

At inference time, the Neural Scheduler operates as a \textit{scoring function} rather than an autoregressive generator.
Unlike autoregressive generation, where token identities are unknown, the tokens of $\mathbf{x}_{\text{out}}$ are fully determined by the onset positions of the voice-line clips: each frame either advances a clip or emits silence. This means that scheduling reduces to a search over onset assignments rather than a generation problem.
As the Neural Scheduler is trained on natural conversation, its NLL reflects how consistent a given timing arrangement is with the conversational dynamics it has observed: arrangements with awkward gaps or poorly timed overlaps will receive high NLL, while natural-sounding ones will not. We therefore use NLL directly as a proxy for timing naturalness. Scheduling then proceeds by searching over candidate onset positions and committing the one that achieves the lowest NLL, effectively asking the model to judge arrangements rather than produce them.
This formulation has three appealing properties: the speech content of every voice line is already fixed at inference time, so there is no risk of drift from the original voice lines; the search is fully deterministic, producing reproducible schedules; and the search space is explicitly parameterized, making the scheduler directly controllable without any modification to model weights, as we demonstrate in \autoref{sec:applications}.

\begin{figure}[t]
\centering
\includegraphics{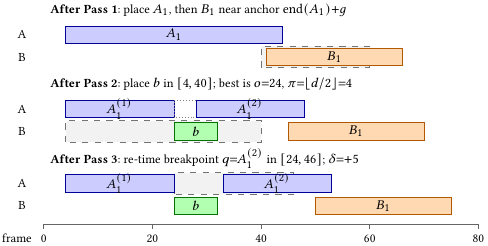}
\caption{An illustrative example of the three passes. \textbf{Pass 1} places the turns, \textbf{Pass 2} inserts the back-channel $b$ with a brief pause in A's turn (dotted), and \textbf{Pass 3} re-times when A resumes. Dashed boxes mark each clip's search range.}
\label{fig:three_pass_example}
\Description{Three stacked two-channel timelines over 80 frames. After Pass 1, turn A1 is followed by the reply B1 with a slight overlap; after Pass 2, a back-channel b is placed under A1, which is split by a short pause; after Pass 3, the second part of A1 and the reply B1 are shifted later.}
\end{figure}

Naively searching over all joint onset assignments is intractable, so we schedule greedily in three sequential passes, each conditioned on the placements committed by the previous one: (1)~primary turns, (2)~back-channels, and (3)~a local re-timing correction. Algorithm~\ref{alg:three_pass} summarizes the procedure, \autoref{fig:three_pass_example} illustrates it on a small example, and we detail each pass below. All timings are expressed in frames (token steps) at the model's $12.5$\,Hz rate.

\paragraph{Scoring a placement}
Given a set of committed onsets, we materialize the corresponding interleaved output region $\mathbf{x}_{\text{out}}$ by advancing the assigned clip on each channel at each frame and emitting $\text{sil}$ elsewhere, then evaluate its NLL under the frozen model conditioned on $\mathbf{x}_{\text{in}}$. Because moving a single clip only perturbs timing locally, we score a window around the clip under consideration, its token span padded by a context margin of $c$ frames on each side, rather than re-scoring the full dialogue; this keeps each candidate evaluation to one cheap forward pass. We write $\mathrm{NLL}(\cdot)$ for this windowed score, where lower is more natural.
We write $\mathcal{S}$ for the set of placements committed so far and $t@o$ for clip $t$ placed with its onset at frame $o$, so $\mathrm{NLL}(\mathcal{S} \cup \{t@o\})$ scores the current schedule with $t$ added at $o$. For a placed clip, $\mathrm{onset}(\cdot)$, $\mathrm{end}(\cdot)$ and $\mathrm{dur}(\cdot)$ give its first frame, end frame and length.

\paragraph{Pass 1: Primary turns}
We first remove all back-channels and place the primary turns in conversational order. Each turn is anchored at the end of the most recently committed turn plus a nominal inter-turn gap $g$ (the first turn is anchored at $g$), and we scan candidate onsets over a window of $\pm W$ frames around the anchor at stride $\sigma$, committing the onset of lowest windowed NLL. A hard floor prevents a turn from overlapping any previously committed turn on the \emph{same} channel, while cross-channel overlap is permitted, which is what allows the model to schedule natural turn-transition overlap. This window is also the primary point of animator control (\autoref{sec:applications}): tightening or shifting the per-speaker bounds forces simultaneous speech or widens inter-turn gaps into deliberate thinking pauses. Turns that contain internal pauses are optionally scheduled sub-segment by sub-segment, searching a short silence between consecutive sub-segments within a per-speaker range.

\paragraph{Pass 2: Back-channels}
Back-channels are placed one at a time, each within the span of its host turn. For a back-channel of duration $d$ hosted by a turn ending at frame $h_{\text{end}}$, we search onsets from the host's start up to $h_{\text{end}} - \lceil r\,d \rceil$, so that at least a fraction $r$ of the back-channel overlaps the host (multiple back-channels on one host are ordered after one another). Rather than an absolute NLL, each candidate is scored by its NLL \emph{difference} $\Delta = \mathrm{NLL}(\text{with back-channel}) - \mathrm{NLL}(\text{baseline without it})$ against the current schedule, isolating the back-channel's own contribution; we commit the minimizing onset. For the first back-channel on a given host we additionally search a host-accommodation strategy jointly with the onset: leave the host unchanged so the back-channel overlaps ongoing speech (\textit{no-pause}), or split the host at the insertion point with a silence gap of $d$ (\textit{full}) or $\lfloor d/2 \rfloor$ (\textit{half}). When a splitting strategy wins, the inserted silence is propagated as a rigid shift to every turn and back-channel scheduled at or after the split point, so downstream timing is preserved.

\paragraph{Pass 3: Local re-timing}
A host split in Pass 2 can leave the remainder of the host turn timed awkwardly relative to the interjection. For each host turn that received a back-channel, we identify \emph{breakpoints}, the sub-segment (or, if the back-channel lands in the host's final segment, the following speaker's turn) immediately after each back-channel. We freeze everything before a breakpoint $q$, search a new onset for it that minimizes NLL, and rigidly shift all subsequent material by the resulting delta; the accumulated shift then propagates to all downstream turns. The search runs from the end of the preceding clip on $q$'s channel up to $\mathrm{onset}(q) + D_q + E$, where $D_q$ is the total duration of the back-channels before $q$ on its host and $E$ bounds the extra silence that may be added, so the resumption can move earlier or later. This mirrors how a speaker naturally adjusts the resumption of their turn after a listener interjects. The scan bounds, gap $g$, overlap ratio $r$, extra gap $E$, and context $c$ are the only free parameters; the exact values used in our experiments, together with the full implementation, are available in our code repository.

\begin{algorithm}[t]
\caption[Three-pass greedy scheduling]{Three-pass greedy scheduling. $\mathcal{S}$ is the schedule built so far and $t@o$ denotes clip $t$ placed at onset frame $o$. Exact search ranges are given in the text, and the full procedure in our code release (\url{https://github.com/EvanSamaa/NeuralSchedularRelease}).}
\label{alg:three_pass}
\begin{algorithmic}[1]
\REQUIRE primary turns $\mathcal{P}$ in script order; back-channels $\mathcal{B}$, each with a host turn; frozen scheduler $p_\theta$
\STATE $\mathcal{S} \gets \emptyset$
\item[] \textit{Pass 1: primary turns}
\FOR{each turn $t \in \mathcal{P}$}
    \STATE $\mathcal{W} \gets$ onsets near the end of the previous turn
    \STATE add $t@o$ to $\mathcal{S}$, with $o = \arg\min_{o' \in \mathcal{W}} \mathrm{NLL}(\mathcal{S} \cup \{t@o'\})$
\ENDFOR
\item[] \textit{Pass 2: back-channels}
\FOR{each $b \in \mathcal{B}$ with host turn $h$}
    \STATE $\mathcal{W} \gets$ onsets that overlap $h$
    \STATE pick $o \in \mathcal{W}$ and a pause for $h$ with lowest NLL
    \STATE add $b@o$ to $\mathcal{S}$; pause $h$ at $o$ and shift later clips
\ENDFOR
\item[] \textit{Pass 3: local re-timing}
\FOR{each clip $q$ that resumes speech after a back-channel}
    \STATE $\mathcal{W} \gets$ onsets shortly after the previous clip
    \STATE move $q$ to its lowest-NLL onset in $\mathcal{W}$; shift later clips
\ENDFOR
\RETURN $\mathbf{x}_{\text{out}}$ rendered from $\mathcal{S}$
\end{algorithmic}
\end{algorithm}

\section{Pipeline Evaluation}
\label{sec:wholistic_evaluation}
To our knowledge, ours is the first computational work to jointly address conversational synthesis through back-channel insertion and turn-clip scheduling. For this reason, we evaluate each component of our framework individually against relevant baselines in \autoref{sec:bc_classifier} and \autoref{sec:neural_scheduler}. We assess the full pipeline in three ways: (1) evaluating back-channel generation and placement using the full-duplex benchmark~\cite{lin2025fullduplexbench}; (2) presenting illustrative conversational examples synthesized via text-to-speech; and (3) conducting a perceptual study to assess the overall system quality.

\subsection{Full-Duplex Bench}
We evaluate the joint performance of the back-channel generation and scheduling pipeline against existing full-duplex dialogue systems using the Full-Duplex Bench evaluation framework~\cite{lin2025fullduplexbench}. The benchmark assesses back-channel behavior along three axes: the Turn-Override Rate (TOR), which measures the proportion of back-channels that exceed a duration threshold and thus intrude on the main speaker's turn; Frequency (Freq), which measures the average rate of back-channel production per second; and the Jensen-Shannon Divergence (JSD) between the predicted back-channel temporal distribution and ground truth, where lower values indicate better alignment with natural back-channel timing.

As shown in Table~\ref{tab:fullduplex}, our proposed system achieves the best performance across all three back-channel metrics. A TOR of 0.000 indicates that none of our generated back-channels override the main speaker; this is trivial, as our architecture only inserts back-channels.
Our system also produces the highest back-channel frequency at 0.062 per second while achieving the lowest JSD of 0.644, suggesting that our model not only is capable of generating more back-channels but also schedules them in a manner most consistent with natural conversational behavior.

\begin{table}[h]
\centering
\caption{Full-Duplex Bench back-channel evaluation results.}
\label{tab:fullduplex}
\begin{tabular}{lccc}
\toprule
& \multicolumn{3}{c}{\textbf{Back-channel}} \\
\cmidrule(lr){2-4}
\textbf{Model} & TOR ($\downarrow$) & Freq ($\uparrow$) & JSD ($\downarrow$) \\
\midrule
Freeze-Omni    & 0.636          & 0.001          & 0.997          \\
Gemini          & 0.091          & 0.012          & 0.896          \\
Moshi           & 1.000          & 0.001          & 0.957          \\
dGSLM           & 0.691          & 0.015          & 0.934          \\
PersonaPlex     & 0.327          & 0.025          & 0.649          \\
\midrule
Proposed (Ours) & \textbf{0.000} & \textbf{0.062} & \textbf{0.644} \\
\bottomrule
\end{tabular}
\Description{Table of turn-override rate, frequency and Jensen-Shannon divergence for back-channels from six systems. Our system has the best value on all three.}
\end{table}
\subsection{Qualitative Examples}
\label{sec:qualitative}

\begin{figure}
    \centering
    \includegraphics[width=0.8\linewidth]{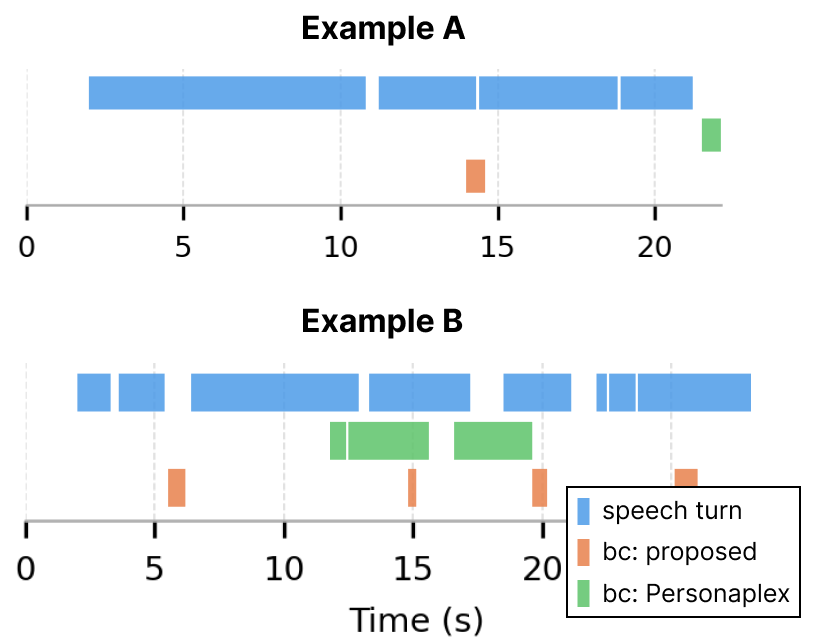}
    \caption{Voice activity timelines for two dialogue examples comparing back-channels generated by PersonaPlex and our proposed model.}
    \label{fig:bc_qualitative}
\Description{Two voice activity timelines, Examples A and B, each showing a speaker's turns together with the back-channels produced by our model and by PersonaPlex. Our back-channels are short and fall within the speaker's turns, while PersonaPlex's are long or come after the speaker has finished.}
\end{figure}

We qualitatively compare our model against baselines on both the back-channel generation and turn scheduling tasks. For back-channel generation, we compare against PersonaPlex~\cite{roypersonaplex}, the state-of-the-art full-duplex audio language model, using examples randomly sampled from the full-duplex benchmark.
\autoref{fig:bc_qualitative} visualizes the voice activity of back-channels inserted by each system. PersonaPlex exhibits two recurring failure modes: either deferring the back-channel until after the speaker has finished (Example A), or taking over the floor during the speaker's turn (Example B). In contrast, our model consistently inserts back-channels either during the speaker's turn or within natural pauses, better reflecting the timing of human listener responses.

\begin{figure}
    \centering
    \includegraphics[width=\linewidth]{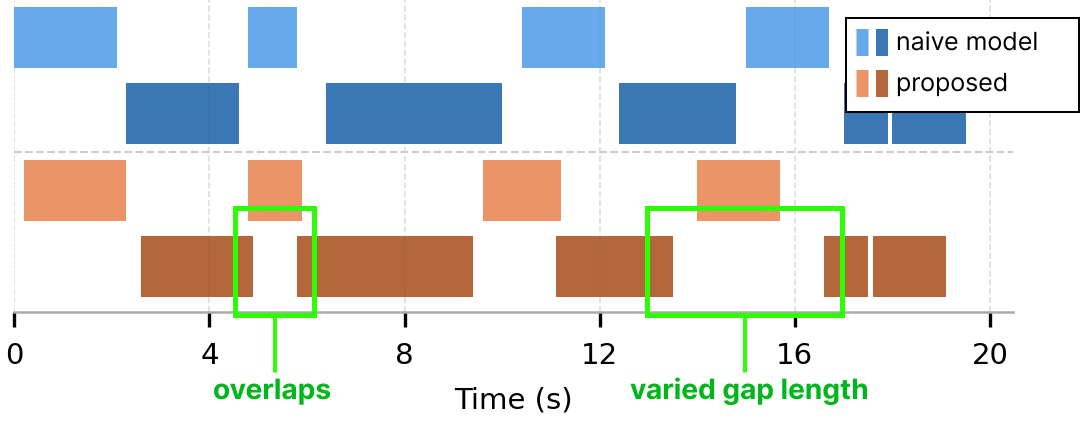}
    \caption{Voice activity timelines for one dialogue, comparing the turn scheduling of the naive pause-based baseline (top) with that of our proposed model (bottom).}
    \label{fig:tt_qualitative}
\Description{Voice activity timeline over 20 seconds comparing a naive fixed-pause baseline (top) with our scheduler (bottom) on the same dialogue. Highlighted regions show that our schedule contains overlapping speech and gaps of varied length, whereas the baseline's gaps are uniform.}
\end{figure}

\paragraph{Turn Scheduling} As no dedicated turn-scheduling baseline exists, we compare against a pause-based heuristic that concatenates voice lines with a fixed inter-turn pause of 200\,ms, an approach used in prior work both for synthesizing two-channel speech datasets~\cite{roypersonaplex} and as an evaluation baseline~\cite{wang2025ntppa, Nguyen2022dGSLM}. We present examples across two settings: casual conversations synthesized with ChatterBox TTS~\cite{chatterboxtts2025}, and more dramatic exchanges sourced from in-the-wild performances from movies and TV shows. As \autoref{fig:tt_qualitative} shows, the naive heuristic produces uniform inter-turn gaps throughout, whereas our scheduler generates varied timing with natural overlaps and pauses, yielding turn-taking dynamics that are markedly less mechanical.

\subsection{User Study}
While our qualitative examples allow readers to assess output quality, we further conduct a forced-choice preference study ($n = 20$ participants) to evaluate perceptual naturalness more rigorously. We showed 10 back-channeling examples, 5 turn-taking examples for conversational speech, and 5 for dramatic speech. In both back-channel and turn-taking conditions, participants listened to pairs of stimuli (ours vs.\ baseline) presented in random order and selected which sounded more natural, with instructions to attend to conversational timing rather than voice quality.

For the back-channel generation examples, our model was significantly preferred over PersonaPlex (67.5\%, 95\% CI [60.7\%, 73.6\%], $p < 0.001$).
Preference for turn scheduling is more nuanced. For casual conversational examples, preference for our model over the naive baseline was at chance ($53.0\%$, 95\% CI $[43.3\%, 62.5\%]$, $p = 0.617$), suggesting that timing differences are difficult to perceive in dialogue with relatively flat turn-taking dynamics. We suspect this reflects the nature of the test stimuli: casual chat and customer service exchanges tend to feature simple, sequential turn structure where the distinction between uniform and natural gaps is subtle. For more dramatic examples featuring real voice-actor recordings, where timing expressivity is more perceptually salient, preference for our model was significant ($64.3\%$, 95\% CI $[52.6\%, 74.5\%]$, $p = 0.022$), suggesting that the scheduler's benefits become perceptually apparent when the source material has richer conversational dynamics. Additional dramatic examples are presented in the supplementary video.

\section{Discussion}
\subsection{Animator Control}
\label{sec:applications}
\begin{figure}[t]
    \includegraphics[width=\linewidth]{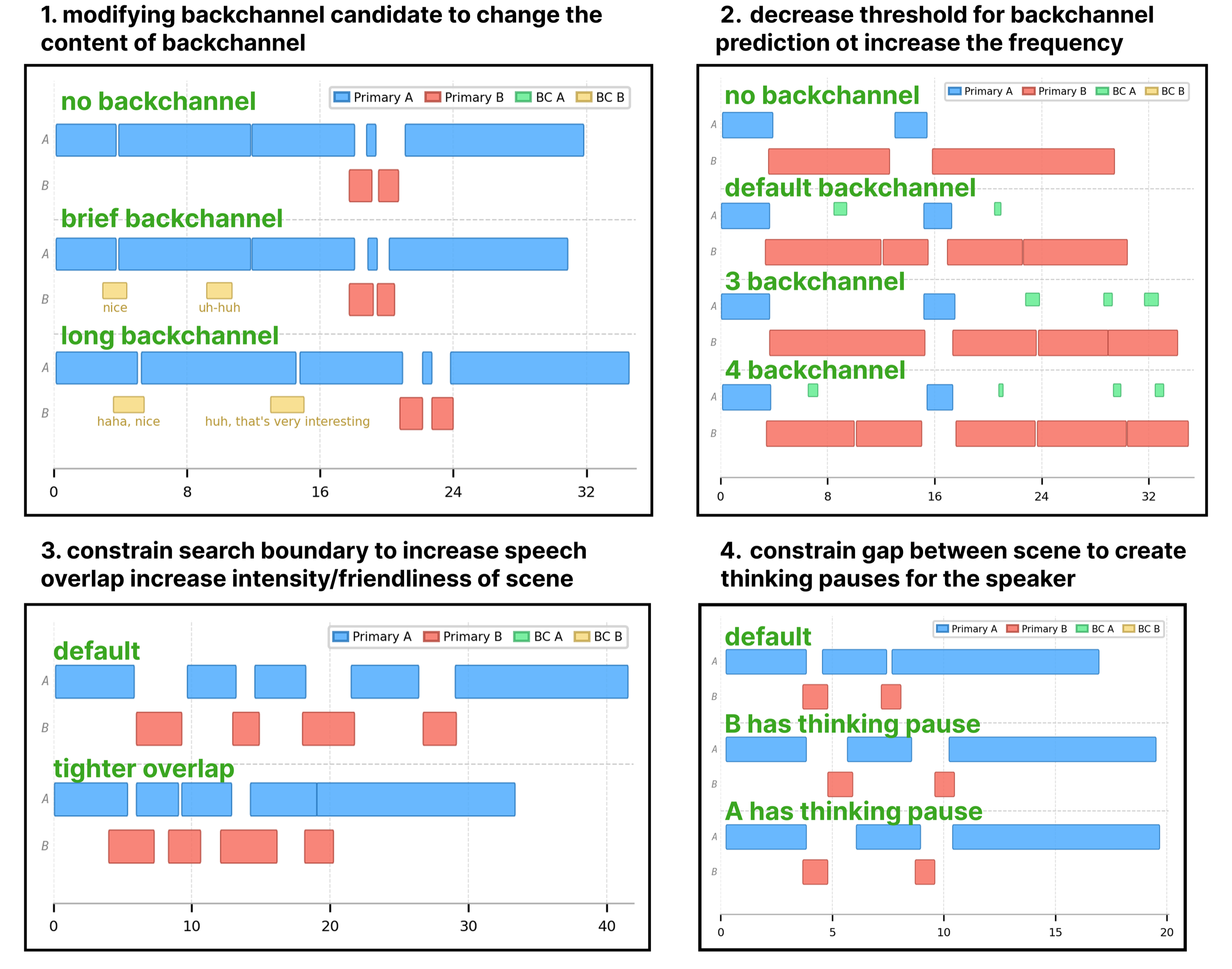}
    \caption{Four axes of animator control: (1)~back-channel content, by swapping candidates; (2)~back-channel frequency, by lowering the prediction threshold; (3)~speech overlap, by tightening onset constraints; and (4)~thinking pauses, by widening one speaker's search window.}
    \label{fig:application}
\Description{Four panels of two-channel voice activity timelines, one per control. Panel 1 swaps back-channel candidates from none to brief to long; panel 2 lowers the threshold to go from zero to four back-channels; panel 3 tightens constraints to increase overlap; panel 4 widens gaps to add thinking pauses for speaker A or B.}
\end{figure}

A key advantage of the search-based inference formulation is that the scheduler's behavior is directly controllable through its search hyperparameters, without modifying model weights. We demonstrate four axes of control in \autoref{fig:application} and the supplementary video. The nature of back-channels can be varied by swapping the back-channel candidates, ranging from short unobtrusive acknowledgements (e.g., ``nice'', ``uh-huh'') to longer, semantically significant responses that cause the primary speaker's turn to be retimed by Pass~3. The number of back-channels is controlled by adjusting the prediction threshold, with lower thresholds producing more frequent back-channels. Overlapping speech can be induced by tightening onset position constraints to force simultaneous activity on both channels. Finally, the search window can be widened for a speaker to introduce natural thinking pauses between turns.

\subsection{Limitations}
While the separation of timing and content is a core strength of our approach, it also limits how back-channels are chosen. The Back-channel Predictor decides whether a clip invites a back-channel and of which class, but not its wording: the back-channel itself is drawn from a vocabulary supplied by the animator. Our system therefore does not produce context-specific back-channels that are semantically grounded in what is being said, unlike voice-LLM-based methods such as PersonaPlex~\cite{roypersonaplex} which jointly model content and timing. This is a property of the Back-channel Predictor rather than of the Neural Scheduler, which simply schedules the clips it is given. The separation is deliberate: it keeps the content of every line, including back-channels, under the animator's direct control, so that the nature and valence of back-channels can be specified by curating the candidate pool. The cost is that this requires manual intervention rather than automatic inference from context. A natural direction for future work is a hybrid approach that uses an LLM to select semantically appropriate back-channel candidates while delegating timing decisions to our scheduler.

A second limitation is that our model operates on fixed voice-line clips and cannot introduce speech disfluencies such as repetitions or stutters, which are also important components of natural conversation. Incorporating disfluency generation into the voice-line synthesis stage can be another promising direction for future work.

\section{Conclusion}

We presented the first controllable neural solution for combining independently recorded voice lines into dynamic conversations with natural turn-taking and back-channeling. By decoupling timing from content, our Neural Scheduler produces conversational dynamics that closely match ground truth while remaining directly controllable: overlap density, back-channel frequency, and character expressiveness can all be specified without any modification to model weights. We believe this separation of timing and content is an unexplored direction for conversational synthesis, with applications ranging from game dialogue production to training data generation for future voice assistants.

\bibliographystyle{ACM-Reference-Format}
\bibliography{references}

\end{document}